\documentclass[prd,reprint,10pt,twocolumn,tightenlines,nofootinbib,showpacs,notitlepage,superscriptaddress,aps]{revtex4-1}
\pdfoutput=1
\usepackage{dcolumn}
\usepackage{amsmath,amsfonts,amssymb,latexsym}
\usepackage[sort&compress]{natbib}
\usepackage{grffile,epstopdf}
\usepackage{graphicx}
\usepackage{hhline}
\usepackage{xcolor}
\usepackage{hhline}
\usepackage{physics}
\usepackage{hyperref}
\usepackage{subcaption}
\usepackage{float}

\makeindex

\begin{document}
	
	\title{Exact relations between the conductivities \\and their connection to the chemical composition of QCD matter}
	
	\author{Jan~A.~Fotakis}
	\email{fotakis@itp.uni-frankfurt.de}
	\affiliation{Institut f\"ur Theoretische Physik, Johann Wolfgang Goethe-Universit\"at,
		Max-von-Laue-Str.\ 1, D-60438 Frankfurt am Main, Germany}
	
	\author{Jakob~E.~Lohr}
	\email{lohr@itp.uni-frankfurt.de}
	\affiliation{Institut f\"ur Theoretische Physik, Johann Wolfgang Goethe-Universit\"at,
		Max-von-Laue-Str.\ 1, D-60438 Frankfurt am Main, Germany}
	
	\author{Carsten Greiner}
	\affiliation{Institut f\"ur Theoretische Physik, Johann Wolfgang Goethe-Universit\"at,
		Max-von-Laue-Str.\ 1, D-60438 Frankfurt am Main, Germany}
	
	\date{\today }
	
	\begin{abstract}
		We present exact relations between the diffusion coefficients or conductivities, $\kappa_{qq'}/T = \sigma_{qq'}$, of strongly-interacting matter. We show that once the diagonal entries are known in two different charge representations, the off-diagonal coefficients are functions of the diagonal entries once isospin symmetry applies. As an important example, we infer the conductivities on the basis of available calculations from lattice quantum chromodynamics (LQCD) and argue that these computations suffer under the approximations made to achieve them. Further, we argue that the representation of the conductivities w.r.t.\,to the conserved quark-flavors may deliver more insight into the chemical composition of strongly-interacting matter.
	\end{abstract}
	
	\maketitle

	\section{Introduction}
	\label{sec:intro}
	
	The primary goal of heavy ion physics is the comprehension of the properties of strongly interacting matter. The collision experiments at particle and heavy-ion accelerators however - which aim to further study these properties - only allow for indirect inferences, which are gained by the employment of theoretical models such as fluid dynamics to describe heavy-ion collision at the current understanding of nature. The underlying fundamental theory in heavy ion physics is quantum chromodynamics (QCD) which describes the strong interaction between quarks and gluons being the fundamental building blocks of nuclear matter. One of the main goals of this field of physics is to eventually understand and to describe the many-body nature of nuclear matter, the chemical composition of that mixture, and the interaction between its components with the help of QCD. In this paper, we focus on the diffusion coefficients, or equivalently, the conductivities of nuclear matter. These describe the response of the strongly-interacting medium to chemical inhomogeneities, or in the case of the electric (cross-)conductivities, to external electromagnetic fields. The investigation of this charge-transporting response may offer another window into the understanding the chemical composition of strongly-interacting matter at extreme temperatures and densities. 
	
	As an important example of that class of phenomena serves the electric conductivity which is directly connected to the lifetime of electromagnetic fields and the generation of the photon- and dilepton spectrum during a heavy-ion collision. Therefore, it has been the object of extensive theoretical investigation, among others, in kinetic theory \cite{Puglisi:2014pda,Puglisi:2014sha,Thakur:2017hfc,Greif:2016skc,Greif:2017byw,Fotakis:2019nbq,Fotakis:2021diq}, various hadronic and partonic transport approaches \cite{Greif:2014oia,Hammelmann:2018ath,Rose:2020sjv}, perturbative QCD \cite{Arnold:2000dr,Arnold:2003zc}, effective field theories \cite{Torres-Rincon:2012sda}, dynamical quasi-particle models \cite{Soloveva:2019xph}, anti-de Sitter/conformal field theory \cite{Finazzo:2013efa,Rougemont:2015ona,Rougemont:2017tlu}, and lattice QCD (LQCD) \cite{Aarts:2007wj,Brandt:2012jc,Amato:2013naa,Aarts:2014nba,Brandt:2015aqk,Ding:2016hua,Astrakhantsev:2019zkr}.
	
	When multiple conserved charges, such as baryon number, electric charge and strangeness, are present in the medium, chemical gradients in one charge may generate currents in the other charges. The magnitude of this coupled-charge transport is not expressed by a single conductivity but a conductivity or diffusion matrix, and it has already been studied for hadronic \cite{Greif:2017byw,Fotakis:2019nbq,Rose:2020sjv,Hammelmann:2023fqw} and partonic matter \cite{Fotakis:2021diq}. Specifically, the off-diagonal components of the conductivity matrix (or also referred to as cross-conductivites) are of interest as they might deliver an insight into chemical composition of nuclear matter as suggested in \cite{Rose:2020sjv}. Such investigations are not yet available from LQCD.
	
	In this paper, we translate the conductivities of strongly-interacting matter from the traditional hadronic charge respresentation into the representation w.r.t.\,the conserved quark-flavor. Since the quarks are the fundamental charge carriers in QCD, we argue that this representation allows for a better understanding of the chemical composition of the medium, which is especially interesting close to the (pseudo-) critical phase transition where the quarks condensate to hadrons. Further, we find relations between the conductivities in both representations, which allows us to evaluate the coefficients from available LQCD results.
	
	This paper is organized as follows: In Section \ref{sec:foundations} we briefly remind the reader of the fundamental definitions and conventions of statistical physics in connection with the diffusion matrix of a system containing conserved quantum numbers. The laws for the transformation between different representations of those conserved charges, obeying the same conservation laws, is derived in Section \ref{sec:transformation_laws}. Section \ref{sec:relations_diffusion_coeff_QCD} provides the explicit transformation laws of the diffusion coefficients of strongly-interacting matter. With the help of these relations, we extract the conductivities from LQCD computations of the isospin conductivity, and we further briefly discuss the approximations made in connection to these results and their viability in Section \ref{sec:extraction}. Finally, in Section \ref{sec:chemical_composition} we sample available results from SMASH, kinetic theory, and LQCD in comparative figures and give them in both the hadronic charge and quark-flavor representation. We further discuss the individual influence of the light- and strange-quark on the conductivities and draw our conclusion in Section \ref{sec:Conclusion}. We make use of natural units $\hbar = c = k_B = 1$.
	
	\section{Foundations}
	\label{sec:foundations}
	
	We consider a system consisting of $N_{\text{spec}}$ different particle species and $N_q$-many conserved intrinsic quantum numbers, which shall be described by the Hamiltonian operator $\hat{H}_s$ and the particle number operators $\hat{N}_s$. In equilibrium, the statistical operator in the grand canonical prescription then reads
	\begin{align}
	\hat{\rho}_G(\beta,\lbrace \alpha_q \rbrace) = \frac{1}{Z_G} \exp\left[ - \sum_{s\,=\,1}^{N_{\text{spec}}} \left( \beta \hat{H}_s + \alpha_s \hat{N}_s \right) \right] \, ,
	\end{align}
	where $\beta = 1/T$ is the inverse temperature, $\alpha_s = \beta \mu_s$ are the specific thermal potentials, and $\mu_s$ is the corresponding specific chemical potential of particle species $s$. In chemical equilibrium, the specific chemical potentials are related to the charge chemical potentials $\mu_{q_i}$ via
	\begin{align}
	\mu_s = \sum_{i\,=\,1}^{N_q} q_{i,s} \, \mu_{q_i} \, ,
	\end{align}
	where $q_{i,s}$ is a conserved intrinsic quantum number (charge) of type $q_i$ of the respective particle species $s$. Here, $i$ enumerates the $N_q$ different conserved charges present in the system, i.e.\,$i \in \lbrace 1, \dots, N_q\rbrace$. Above, we further defined the grand canonical partition function as
	\begin{align}
	Z_G \equiv \tr( \exp\left[ - \sum_{s\,=\,1}^{N_{\text{spec}}} \left( \beta \hat{H}_s + \alpha_s \hat{N}_s \right) \right] ) \, .
	\end{align}
	The ensemble average of an arbitrary operator $\hat{A}$ is then defined as the trace of the left-product of the statistical operator with $\hat{A}$ via
	\begin{align}
	\langle \hat{A} \rangle &= \tr( \hat{\rho}_G \hat{A} ) \, .
	\end{align}
	If $N^\mu_s$ is the ensemble-averaged four-current of the given particle species, the information about the transported charge of given type $q_i$ is contained in the corresponding charge four-current defined by
	\begin{align}
	N^\mu_{q_i} \equiv \sum_{s\,=\,1}^{N_{\text{spec}}} q_{i,s} N^\mu_s = n_{q_i} u^\mu + V^\mu_{q_i} \, .
	\end{align}
	If the system just consisted of fermions, the specific four-current of each species is $N^\mu_s \equiv \langle \bar{\psi}_s \gamma^\mu \psi_s \rangle$. Here, $\psi_s$ is the respective spinor field, $\gamma^\mu$ are the Dirac operators, and $\bar{\psi}_s = \psi_s^\dagger \gamma^0$ is the Dirac adjoint to $\psi_s$. Above, we have further given the decomposition of the charge current with respect to the local fluid or collective velocity $u^\mu$, and by defining the projector $\Delta^{\mu\nu} = g^{\mu\nu} - u^\mu u^\nu$ we can then express the (space-like) diffusion current by the projection of the charge four-current, i.e.\,$V^\mu_{q_i} = \Delta^\mu_{~\nu} N^\nu_{q_i}$.
	
	Once the medium experiences a small, external gradient in the thermal potentials $\alpha_{q_i} = \mu_{q_i}/T$ according to Navier-Stokes theory \cite{DeGroot:1980dk}, and thus the resulting linear response of the system then obeys
	\begin{align}
	V^\mu_{q_i} = \sum_{j\,=\,1}^{N_q} \kappa_{q_i q_j} \nabla^\mu \alpha_{q_j} \, . \label{eq:navier_stokes}
	\end{align}
	Here, $\nabla^\mu = \Delta^{\mu\nu} \partial_\nu$ is the orthogonal, spatial gradient in flat Minkowski spacetime, and $\kappa_{q_i q_j}$ are the so-called diffusion coefficients which characterize the magnitude of the system response. The diffusion coefficients are directly related to the associated conductivities via $\sigma_{q_iq_j} = \kappa_{q_iq_j}/T$, and thus we use both terms and symbols interchangeably in this paper for convenience. 
	
	So far we have not yet specified the exact nature of the conserved intrinsic quantum numbers and their representation. In the following we investigate how the transport coefficients of any system behave under transformations from one chosen charge representation to the other.
	
	\section{Transformations in charge representation space}
	\label{sec:transformation_laws}
	
	Let $\boldsymbol{q}_s = ( q_{1,s},\dots, q_{N_q,s} )$ be the vector of conserved intrinsic quantum numbers carried by a particle of species $s$. We can choose a different representation of the conserved charges, $\boldsymbol{q}'_s = ( q'_{1,s},\dots, q'_{N_q,s} )$, described by the transformation $\boldsymbol{q}_s \mapsto \boldsymbol{q}'_s = \boldsymbol{q}'_s(\boldsymbol{q}_s)$. Under a valid transformation, the number of conservation laws remains the same and the charge vectors in both representations are thus of equal dimensionality. The transformation in charge representation space then is prescribed by a $(N_q \times N_q)$-matrix $\hat{\mathcal{M}}$ of full rank with components $\mathcal{M}_{ij}$ via
	\begin{align}
	\boldsymbol{q}'_s = \hat{\mathcal{M}}\, \boldsymbol{q}_s \quad \text{or} \quad q'_{i,s} = \sum_{j\,=\,1}^{N_q} \mathcal{M}_{ij} \, q_{j,s} \, . \label{eq:transform}
	\end{align}
	And since $\hat{\mathcal{M}}$ shall represent a valid transformation, the matrix is regular and the back-transformation reads
	\begin{align}
	\boldsymbol{q}_s = \hat{\mathcal{M}}^{-1}\, \boldsymbol{q}'_s \, . \label{eq:back_transform}
	\end{align}
	From now on, the bold-printed quantities represent a vector in charge representation space. For clarity we write down the vector of charge chemical potentials and the vector of charge currents as important examples:
	\begin{align}
	\boldsymbol{\alpha}_q &\equiv ( \alpha_{q_1}, \dots, \alpha_{q_{N_q}} ) \, , \\
	\boldsymbol{N}^\mu_q &\equiv ( N^\mu_{q_1}, \dots, N^\mu_{q_{N_q}} ) \, .
	\end{align}
	The grand canonical partition function, especially the specific thermal potentials $\alpha_s$, do not depend on the explicit representation in charge space and thus are invariant under such transformations. We therefore have the following condition in matrix representation
	\begin{align}
	\alpha_s = \boldsymbol{q}_s \cdot \boldsymbol{\alpha}_q = \boldsymbol{q}'_s \cdot \boldsymbol{\alpha}_{q'} \, , \label{eq:cond_chem_pot}
	\end{align}
	from which we retrieve the well-known transformation law for the charge chemcial potentials after applying Eq.\,\eqref{eq:transform}:
	\begin{align}
	\boldsymbol{\mu}_{q'} = \left( \hat{\mathcal{M}}^{-1} \right)^T \boldsymbol{\mu}_q \, . \label{eq:TL_chemical_pot}
	\end{align}
	All quantities which are directly proportional to charge transform equally as does the charge itself. Therefore, the transformation laws of the charge and diffusion currents read
	\begin{align}
	\boldsymbol{N}^\mu_{q'} = \hat{\mathcal{M}} \, \boldsymbol{N}^\mu_q \, , \\
	\boldsymbol{V}^\mu_{q'} = \hat{\mathcal{M}} \, \boldsymbol{V}^\mu_q \, .
	\end{align}
	Applying the above relations to the Navier-Stokes law \eqref{eq:navier_stokes} for the diffusion currents in the form,
	\begin{align}
	\boldsymbol{V}^\mu_{q} = \hat{\kappa} \,\nabla^\mu \boldsymbol{\alpha}_q \, ,
	\end{align} 
	directly leads to a transformation law of the conductivities:
	\begin{align}
	\hat{\kappa}' = \hat{\mathcal{M}}\, \hat{\kappa} \, \hat{\mathcal{M}}^T \, . \label{eq:TL_diffusion_coeff}
	\end{align}
	With the transformation laws above, one can directly derive such rules for all transport coefficients derived in Refs.~\cite{Monnai:2010qp,Kikuchi:2015swa,Fotakis:2022usk,Hu:2022vph}. For this we refer to Ref.\,\cite{Fotakis:2022usk}. There are two families of transport coefficients: one, like shear viscosity or bulk viscosity, which do not dependent on charge, and the others, like the conductivities, which do. In the family of charge-dependent transport coefficients there is the set of coefficients which are direct proportional to $q_{i,s}$ (see Eqs.\,(C3), (C4), (C8) - (C11), (C17) and (C18) in Ref.\,\cite{Fotakis:2022usk}) and therefore transform identically. The remaining coefficients are directly proportional to products of the form $q_{i,s} q_{j,s}$ (see Eqs.\,(116), (117), (C5) - (C7), (C12), (C13), and (C19) in Ref.\,\cite{Fotakis:2022usk}). The main object of the latter is the diffusion coefficient matrix which, according to Eq.\,(116) in Ref.\,\cite{Fotakis:2022usk}, can be written as
	\begin{align}
	\hat{\kappa} = \sum_{s,s'\,=\,1}^{N_{\text{spec}}} \boldsymbol{q}_s \boldsymbol{q}^T_{s'} \tilde{\kappa}_{ss'} \- , \label{eq:diff_matrix}
	\end{align}
	Here, we defined
	\begin{align}
	\tilde{\kappa}_{ss'} \equiv \sum_{\tilde{s}\,=\,1}^{N_{\text{spec}}} \sum_{r\,=\,0}^{N_1} \tau^{(1)}_{s\tilde{s},0r} \left( \delta_{\tilde{s}s'} J_{\tilde{s},r+1,1} - \frac{n_{s'}}{e + P_0} J_{\tilde{s},r+2,1} \right) \, ,
	\end{align}
	which does not depend on the chosen charge representation. In that definition, $\tau^{(1)}_{s\tilde{s},r'r}$ contains the relaxation times of the system response, and $J_{\tilde{s},r,q}$ are the auxillary thermodynamic functions to species $\tilde{s}$. See Ref.\,\cite{Fotakis:2022usk} for details. Provided that time reversal symmetry of the underlying interactions is a given, we first note the symmetry of $\hat{\kappa}$, which is well-known from the Onsager relations\cite{Onsager1931a,Onsager1931b}:
	\begin{align}
	\hat{\kappa}^T = \left( \sum_{s,s'\,=\,1}^{N_{\text{spec}}} \boldsymbol{q}_s \boldsymbol{q}^T_{s'} \tilde{\kappa}_{ss'} \right)^T = \sum_{s,s'\,=\,1}^{N_{\text{spec}}} \boldsymbol{q}_{s'} \boldsymbol{q}^T_{s} \tilde{\kappa}_{ss'}  = \hat{\kappa} \, ,
	\end{align}
	where in the last step we relabeled $s \leftrightarrow s'$  and used that under time reversal symmetry it is $\tilde{\kappa}_{ss'} = \tilde{\kappa}_{s's}$. We further directly recover the transformation law from Eq.\,\eqref{eq:TL_diffusion_coeff} since we have
	\begin{align}
	\hat{\kappa}' &= \sum_{s,s'\,=\,1}^{N_{\text{spec}}} \boldsymbol{q}'_s \boldsymbol{q}'^T_{s'} \tilde{\kappa}_{ss'} = \sum_{s,s'\,=\,1}^{N_{\text{spec}}} \hat{\mathcal{M}} \boldsymbol{q}_s \boldsymbol{q}^T_{s'} \hat{\mathcal{M}}^T \tilde{\kappa}_{ss'} \\
	&= \hat{\mathcal{M}}\, \hat{\kappa}\, \hat{\mathcal{M}}^T \, . \notag
	\end{align}
	One can diagonalize the conductivity matrix by defining $\hat{\mathcal{P}} = (\boldsymbol{v}_1, \dots, \boldsymbol{v}_{N_q} )$, the matrix which contains the eigenvectors of $\hat{\kappa}$, and writing
	\begin{align}
	\hat{\kappa}_{\text{diag}} = \hat{\mathcal{P}}^{-1} \hat{\kappa} \hat{\mathcal{P}} \, .
	\end{align}
	Since $\hat{\kappa}$ is real and symmetric, its eigenvectors can thus be chosen to be orthonormal, $\boldsymbol{v}_i \cdot \boldsymbol{v}_j = \delta_{ij}$, and $\mathcal{P}$ is thus unitary, $\mathcal{P}^{-1} = \mathcal{P}^T$. Therefore, $\mathcal{P}^T$ also defines a transformation in charge representation space. However, there are cases where the conductivity matrix is already diagonal without imposing above transformations. In good approximation, this is e.g.\,the case in the quark-flavor representation (see next Section) of an ultrarelativistic quark-gluon plasma (QGP) where it is believed that only quasi-free quarks and gluons and their anti-particles are the degrees of freedom. Here, the off-diagonal terms of the linear combinations of $\boldsymbol{q}_s \boldsymbol{q}^T_{s'}$ cancel due to the presence of the anti-partner of each species, provided that the masses and the cross sections are the same for all particles.
	
	Most second-order coefficients, which are quadratically proportional to charge, fulfill similar relations as discussed above. In the following however, we will restrict ourselves to the discussion of the conductivities in the case of strongly-interaction matter.	
	
	\section{Relations between the conductivities of QCD matter}
	\label{sec:relations_diffusion_coeff_QCD}
	
	The above given transformation law of the conductivity matrix generally allow for expression of all off-diagonal coefficients in terms of the diagonal entries. In this section we discuss such relations for strongly-interacting matter, yet this approach can also be applied to any given physical system with conserved charge.
	
	According to quantum chromodynamics (QCD), the fundamental charge-carriers of strongly-interacting matter are the quarks. In the case of $N_f = 2+1$ flavors, the conserved charges are the quark-flavors up- ($u$), down- ($d$), and strange-flavor ($s$). In the hadronic-charge representation, the conservation of quark-flavor translates into the conservation of baryon number ($B$), electric charge ($Q$), and strangeness ($S$). If we enumerate the quarks with $s' = \lbrace 1,2,3 \rbrace$ (in the order up-, down-, and strange-quark), the charge vectors in the two representations read as follows. In the quark-flavor ($uds$) representation, $\boldsymbol{f}_{s'} = (u_{s'}, d_{s'}, s_{s'})$, we have
	\begin{align}
	\boldsymbol{f}_1 = \begin{pmatrix} 1 \\ 0 \\ 0 \end{pmatrix}\, , ~ \boldsymbol{f}_2 = \begin{pmatrix} 0 \\ 1 \\ 0 \end{pmatrix}\, , ~ \boldsymbol{f}_3 = \begin{pmatrix} 0 \\ 0 \\ 1 \end{pmatrix}\, .
	\end{align}
	In the hadronic charge ($BQS$) representation, $\boldsymbol{c}_{s'} = (B_{s'}, Q_{s'}, S_{s'})$, we have
	\begin{align}
	\boldsymbol{c}_1 = \begin{pmatrix} 1/3 \\ 2e/3 \\ 0 \end{pmatrix}\, , ~ \boldsymbol{c}_2 = \begin{pmatrix} 1/3 \\ -e/3 \\ 0 \end{pmatrix}\, , ~ \boldsymbol{c}_3 = \begin{pmatrix} 1/3 \\ -e/3 \\ -1 \end{pmatrix}\, .
	\end{align}
	Here, $e \approx \sqrt{4\pi/137}$ is the unit electric charge in natural units. The transformation from $uds$- to $BQS$-representation, $\boldsymbol{c}_s = \hat{\mathcal{M}} \boldsymbol{f}_s$, then is given by the following matrix
	\begin{align}
	\hat{\mathcal{M}} = 
	\begin{pmatrix}
	1/3 & 1/3 & 1/3 \\ 
	2e/3 & -e/3 & -e/3 \\
	0 & 0 & -1
	\end{pmatrix} \, .
	\end{align}
	The back-tranformation reads
	\begin{align}
	\hat{\mathcal{M}}^{-1} = 
	\begin{pmatrix}
	1 & 1/e & 0 \\ 
	2 & -1/e & 1 \\
	0 & 0 & -1
	\end{pmatrix} \, .
	\end{align}
	The chemical potentials then transform according to Eq.~\eqref{eq:TL_chemical_pot} and explicitly read
	\begin{align}
	\mu_B &= \mu_u + 2\mu_d \, , \notag \\ 
	e\,\mu_Q &= \mu_u - \mu_d \, , \notag \\ 
	\mu_S &= \mu_d - \mu_s \, ,
	\end{align}
	and
	\begin{align}
	\mu_u &= \frac{1}{3} \mu_B + \frac{2}{3} e\, \mu_Q\,, \notag \\ 
	\mu_d &= \frac{1}{3} \mu_B - \frac{e}{3} \mu_Q\, , \notag \\ 
	\mu_s &= \frac{1}{3} \mu_B - \frac{e}{3} \mu_Q - \mu_S \, ,
	\end{align}
	which are also well-known. The coefficients transform via Eq.~\eqref{eq:TL_diffusion_coeff}, and explicitly read
	\begin{align}
	\kappa_{BB} &= \frac{1}{9} \big( \kappa_{uu} + \kappa_{dd} + \kappa_{ss} + 2\kappa_{ud} + 2\kappa_{us} + 2\kappa_{ds} \big) \, , \notag \\
	\kappa_{QQ} &= \frac{e^2}{9} \big( 4\kappa_{uu} + \kappa_{dd} + \kappa_{ss} + 2\kappa_{ds} - 4\kappa_{ud} - 4\kappa_{us} \big) \, , \notag \\
	\kappa_{SS} &= \kappa_{ss} \, , \notag \\
	\kappa_{BQ} &= \frac{e}{9} \big( 2\kappa_{uu} - \kappa_{dd} - \kappa_{ss} + \kappa_{ud} + \kappa_{us} - 2\kappa_{ds} \big) \, , \notag \\
	\kappa_{BS} &= -\frac{1}{3} \big( \kappa_{ss} + \kappa_{us} + \kappa_{ds} \big) \, , \notag \\
	\kappa_{QS} &= \frac{e}{3} \big( \kappa_{ss} + \kappa_{ds} - 2\kappa_{us} \big) \, , \label{eq:relations_diff_coeff_QCD}
	\end{align}
	and the back-transformation is
	\begin{align}
	\kappa_{uu} &= \kappa_{BB} + \frac{1}{e^2} \kappa_{QQ} + \frac{2}{e} \kappa_{BQ} \, , \notag \\
	\kappa_{dd} &= 4\kappa_{BB} + \frac{1}{e^2} \kappa_{QQ} + \kappa_{SS} - \frac{4}{e}\kappa_{BQ} + 4\kappa_{BS} - \frac{2}{e}\kappa_{QS} \, , \notag \\
	\kappa_{ss} &= \kappa_{SS} \, , \notag \\ 
	\kappa_{ud} &= 2\kappa_{BB} - \frac{1}{e^2} \kappa_{QQ} + \frac{1}{e} \kappa_{BQ} + \kappa_{BS} + \frac{1}{e}\kappa_{QS} \, , \notag \\
	\kappa_{us} &= - \left( \kappa_{BS} + \frac{1}{e}\kappa_{QS} \right) \, , \notag \\
	\kappa_{ds} &= - \kappa_{SS} - 2\kappa_{BS} + \frac{1}{e} \kappa_{QS} \, .
	\end{align}
	Further, we reiterate that the diffusion coefficient matrix is symmetric, i.e.\,$\kappa_{qq'} = \kappa_{q'q}$ \cite{Onsager1931a,Onsager1931b}. 
	
	Since the quark-flavor charge coincides with the corresponding quark count in QCD, the plausibility of the above relations is directly verifiable. For each of the $BQS$-conductivities we directly recognize the explicit form of Eq.\,\eqref{eq:diff_matrix}, where the prefactors simply are the assigned hadronic charge of each contributing quark-quark combination, i.e.\newline$\kappa_{BS} = \sum\limits_{f,f'\,=\,u,d,s} B_{f} S_{f'} \kappa_{ff'} = -(\kappa_{ss} + \kappa_{us} + \kappa_{ds})/3$. The relations of the quark-flavor chemical potentials also reflect the charge of each quark, e.g.\,$\mu_u = \frac{1}{3}\mu_B + \frac{2}{3}e \, \mu_Q$. As an illustration of the connection between the quark-flavor and hadronic chemical potentials, we could think of the proton which consists of two up-quarks and one down-quark. Then the proton chemical potential is $\mu_p = 2\mu_u + \mu_d = \mu_B + e\,\mu_Q$ (see also Eq.\,\eqref{eq:cond_chem_pot}), since the proton carries one unit each of the electric and baryon charge.
	
	From above it is clear that, due to $\kappa_{ss} = \kappa_{SS}$, in this choice of representations it is not possible to express all off-diagonal coefficients by diagonal ones. Thus, once the diagonal coefficients in both representations ($\kappa_{BB}$, $\kappa_{QQ}$, $\kappa_{ss}$, $\kappa_{uu}$, $\kappa_{dd}$) \textit{and} one other off-diagonal entry (here $\kappa_{QS}$) are known, we can express the remaining (off-diagonal) coefficients explicitly with the help of Eqs.~\eqref{eq:relations_diff_coeff_QCD}. For this we define
	\begin{align}
	\Lambda_1 &\equiv 9\kappa_{BB} - \left( \kappa_{uu} + \kappa_{dd} + \kappa_{ss} \right) \, , \notag \\
	\Lambda_2 &\equiv -\frac{9}{e^2}\kappa_{QQ} + 4\kappa_{uu} + \kappa_{dd} + \kappa_{ss} \, , \notag \\
	\Lambda_3 &\equiv 2\kappa_{uu} - \kappa_{dd} - \kappa_{ss} \, ,
	\end{align}
	where $\Lambda_i$ are functions exclusively of the diagonal entries, and find
	\begin{align}
	\kappa_{ud} &= \frac{1}{4}\Lambda_2 - \frac{1}{2}\kappa_{ss} + \frac{3}{2e} \kappa_{QS} \, , \notag \\
	\kappa_{us} &= \frac{1}{12}\left( 2\Lambda_1 - \Lambda_2 + 6\kappa_{ss} \right) - \frac{3}{2e}\kappa_{QS} \, , \notag \\
	\kappa_{ds} &= \frac{1}{6} \left( 2\Lambda_1 - \Lambda_2 \right) \, , \notag \\
	\kappa_{BQ} &= \frac{e}{9}\left[ \Lambda_3 + \frac{1}{2}\left( \Lambda_2 - \Lambda_1 \right) \right] \, , \notag \\
	-\kappa_{BS} &= \frac{1}{12}\left( 2\Lambda_1 - \Lambda_2 + 6\kappa_{ss} \right) - \frac{1}{2e} \kappa_{QS} \, . \label{eq:off_diag_kappa_to_diag}
	\end{align}
	We remind the reader that we further have
	\begin{align}
	\kappa_{QS} &= \frac{e}{3} \big( \kappa_{ss} + \kappa_{ds} - 2\kappa_{us} \big) \, .
	\end{align}
	Thus, under the assumption of isospin symmetry, where the conductivities of the light-sector ($\ell$) obey the equalities $\kappa_{uu} = \kappa_{dd} \equiv \kappa_{\ell\ell}$ and $\kappa_{us} = \kappa_{ds} \equiv \kappa_{\ell s}$, it is $\kappa_{QS} = \frac{e}{3} \big( \kappa_{ss} - \kappa_{\ell s} \big)$ and the above expressions reduce to
	\begin{align}
	\kappa_{ud} &= \frac{1}{6} \left( 2\Lambda_2 - \Lambda_1 \right) \, , \notag \\
	\kappa_{\ell s} &= \frac{1}{6} \left( 2\Lambda_1 - \Lambda_2 \right) = -\kappa_{ud} \, , \notag \\
	\kappa_{BQ} &= \frac{e}{9}\left[ \Lambda_3 + \frac{1}{2}\left( \Lambda_2 - \Lambda_1 \right) \right] \, , \notag \\
	-\kappa_{BS} &= \frac{1}{3}\kappa_{ss} + \frac{1}{9} \left( 2\Lambda_1 - \Lambda_2 \right) \, . \label{eq:off_diag_kappa_to_diag_isopin_symm}
	\end{align}
	Note that in the case of $N_f = 2$ flavors (light quarks only), the coefficients connected to the strange-quark vanish, i.e.\,$\kappa_{ss} = 0$ and $\kappa_{\ell s} = 0$, and above relations change accordingly.
	
	\section{Extraction of all conductivities from lattice QCD results}
	\label{sec:extraction}
	
	In the recent history of publications of results from Lattice QCD, a set of crucial approximations was assumed in order to calculate the conductivities. In simplified terms, the coefficients $\kappa_{qq'}$ are directly related to the zero-momentum limit of the quark-spectral functions, which in turn are extracted from the Euclidean vector current correlator. The latter is defined as the ensemble averaged charge four-current correlation between currents of charge type $q$ and those of charge type $q'$ as \cite{Aarts:2007wj,Brandt:2012jc,Amato:2013naa,Aarts:2014nba,Brandt:2015aqk,Ding:2016hua,Astrakhantsev:2019zkr}
	\begin{align}
	G^{\mu\nu}_{qq'}(\tau) &= \int \dd^3 \mathbf{x} \expval{ N^\mu_q(\tau,\mathbf{x}) N^\nu_{q'}(0,\mathbf{0})^\dagger } \\
	&= \sum_{s,s'\,=\,1}^{N_{\text{spec}}} q_s q'_{s'} \int \dd^3 \mathbf{x} \expval{ N^\mu_s(\tau,\mathbf{x}) N^\mu_{s'}(0,\mathbf{0})^\dagger } \, . \notag
	\end{align}
	The calculation then is carried out at zero chemical potential ($\mu_q = 0$), and it is assumed that if the quarks are degenerate, the off-diagonal quark-quark current correlation ($s \neq s'$) can be neglected \cite{Aarts:2007wj,Brandt:2012jc,Amato:2013naa,Aarts:2014nba,Brandt:2015aqk,Ding:2016hua,Astrakhantsev:2019zkr}, and thus
	\begin{align}
	G^{\mu\nu}_{qq'}(\tau) \approx \sum_{s\,=\,1}^{N_{\text{spec}}} q_s q'_s \int \dd^3 \mathbf{x} \expval{ N^\mu_s(\tau,\mathbf{x}) N^\mu_s(0,\mathbf{0})^\dagger } \, .
	\end{align}
	This would be a legitimate approximation at high temperatures when the asymptotic limit is approached, where the quarks and gluons are believed to be quasi-free and uncorrelated. Further, it may be assumed that the quarks contribute equally, e.g.\,in the case of $N_f = 2$ isospin symmetry is taken into account. Then, the above formula finally becomes
	\begin{align}
	G^{\mu\nu}_{qq'}(\tau) \approx G^{\mu\nu}_{\text{quark}} \sum_{s\,=\,1}^{N_{\text{spec}}} q_s q'_s  \, , \label{eq:lattice_approx}
	\end{align}
	where
	\begin{align}
	G^{\mu\nu}_{\text{quark}} \equiv \int \dd^3 \mathbf{x} \expval{ N^\mu_{\text{quark}}(\tau,\mathbf{x}) N^\nu_{\text{quark}}(0,\mathbf{0})^\dagger }
	\end{align}
	is the current-current correlator of any type of quark irrespective of its flavor. For $N_f = 2$ this is sometimes referred to as the isospin current correlator. Under these assumptions, even the diagonal conductivities are now related to each other. We have $\kappa_{BB} = C_B \kappa_{\ell\ell}$, $\kappa_{QQ} = C_{\text{el}} \kappa_{\ell\ell}$, and if we assume that the strange-quark contributes equally like a light quark, it also is $\kappa_{SS} = \kappa_{ss} = C_S \kappa_{\ell\ell}$. Using Eqs.~\eqref{eq:off_diag_kappa_to_diag_isopin_symm}, the problem of evaluating the complete conductivity matrix is reduced to the extraction of the light-light quark conductivity $\kappa_{\ell\ell}$ from the lattice.
	
	Above, we defined the coefficients
	\begin{align}
	C_B \equiv \sum_i (B_i)^2 \, , ~ C_{\text{el}} \equiv \sum_i (Q_i)^2 \, , ~ C_S \equiv \sum\limits_i (S_i)^2 \, ,
	\end{align}
	among which $C_{\text{el}}$ is prominent \cite{Aarts:2007wj,Brandt:2012jc,Amato:2013naa,Aarts:2014nba,Brandt:2015aqk,Ding:2016hua,Astrakhantsev:2019zkr}.
	In the case of $N_f = 2$ these become $C_B = \frac{2}{9}$, $C_{\text{el}} = \frac{5}{9}e^2$, and $C_S = 0$. From Eqs.~\eqref{eq:off_diag_kappa_to_diag_isopin_symm} it then follows that
	\begin{align}
	&\kappa_{\ell s} = 0 \, , \quad \kappa_{ud} = 0 \, , \notag \\ 
	&\kappa_{BQ} = \frac{e}{9} \kappa_{\ell\ell} \, , \quad \kappa_{BS} = 0 \, , \quad \kappa_{QS} = 0 \, .
	\end{align}
	For $N_f = 3$ the coefficients then evaluate to $C_B = \frac{1}{3}$, $C_{\text{el}} = \frac{2}{3}e^2$, and $C_S = 1$. The off-diagonal conductivities now read 
	\begin{align}
	&\kappa_{\ell s} = 0 \, , \quad \kappa_{ud} = 0 \, , \notag \\ 
	&\kappa_{BQ} = 0 \, , \quad \kappa_{BS} = -\frac{1}{3} \kappa_{\ell\ell} \, , \quad \kappa_{QS} = \frac{e}{3} \kappa_{\ell\ell} \, . \label{eq:lattice_QCD_Nf2+1}
	\end{align}
	With these relations, one can extract all conductivities from available results from lattice QCD.
	
	It is important to note that the off-diagonal quark-flavor conductivities, $\kappa_{ud}$ and $\kappa_{\ell s}$, always vanish due to the negligence of the off-diagonal quark-quark current correlation. However, from hadronic calculations it is expected that these are non-zero in the vicinity of the pseudo-critical temperature. Thus the extracted results from lattice QCD calculations deviate from this expectation. The conductivities may provide some insight regarding the chemical composition of the strongly-interacting matter. With this in mind,it would be interesting to see these approximations relaxed in future calculations on the lattice.
	
	\section{Insights into the chemical composition of strongly-interacting matter}
	\label{sec:chemical_composition}
	
	The conductivities ``regulate'' the magnitude of the charge flow in response to an external chemical gradient. Therefore, it was already motivated in Ref.~\cite{Rose:2020sjv} that they may deliver some indications regarding the chemical composition of the flowing medium. For example, the cross-conductivity $\sigma_{QB}$ relates to the electric flow resulting from a baryo-chemical gradient. This coefficient is only non-zero if either there are (temporary) bound states in the medium which carry both charge types and are thus sensitive to these gradients, or if there is sufficient interaction between flowing baryons and electric charge such that the baryonic particles drag the electric particles with them. Further, the magnitude of the conductivities grows proportionally with the corresponding charge content while it is suppressed by the scattering rate of the participating constituents \cite{Fotakis:2019nbq,Fotakis:2021diq}. It should be mentioned that light particles dominate the contribution to these coefficients since heavy particles are thermally suppressed. In other words, The coefficients are directly related to the correlation of flowing charges and their distribution on the constituents of the medium.
	
	In QCD the quark-flavor representation ($uds$) of the transport coefficients may paint a clearer picture on that chemical composition in terms of quarks than the hadronic charge representation ($BQS$). This is because in the asymptotic limit, where the quarks and gluons are quasi-free and thus are (weakly or) uncorrelated, the conductivity matrix in the $uds$-representation is diagonal while in the $BQS$-representation - which is a linear combination of the quark-flavors - it is not. Therefore, deviations from this expectation may lead to inferences on the abundance of quark-composite states e.g.\,due to their condensation to hadrons or due to the presence of hadronic excitations at higher temperatures\cite{Bala:2023iqu}. In Figure \ref{fig:kappa_BQS_Nf2+1_BQS} we show a collection of available results to the full conductivity matrix in the $BQS$-representation from both: Hadronic and partonic approaches. While in Figure \ref{fig:kappa_uds_Nf2+1_uds} we provide the same collection in the $uds$-representation. On the hadronic side, results from the hadronic transport model SMASH \cite{SMASH:2016zqf} were recently published\cite{Hammelmann:2023fqw}. There are also results available from the first-order Chapman-Enskog (CE) method for a hadron gas (denoted as HRG) consisting of the lightest hadronic species - namely a gas consisting of pions, kaons, nucleons, $\Lambda$- and $\Sigma$-baryons \cite{Greif:2017byw,Fotakis:2019nbq}. Moreover, there exist results for a partonic system consisting of quarks and gluons with the help of the dynamical quasi-particle model (DQPM) \cite{Soloveva:2019xph}, which were achieved from the CE method \cite{Fotakis:2021diq}. Furthermore, results based on lattice QCD calculations are presented, which were attained on grounds of Eqs.\,\eqref{eq:off_diag_kappa_to_diag_isopin_symm} together with the given LQCD result for $\sigma_{\ell\ell} = C^{-1}_{\text{el}} \sigma_{QQ}$ \cite{Aarts:2014nba,Brandt:2015aqk,Astrakhantsev:2019zkr}. Here, we made the assumption that we have up-, down-, and strange-quarks, which contribute equally to the coefficients ($N_f = 3$), i.e.\,$C_B = 1/3$, $C_{\text{el}} = 2e^2/3$, and $C_s = 1$. The comparison is depicted as a function of $T/T_c$ for vanishing chemical potentials, $\mu_q = 0$, where $T_c$ is the pseudo critical temperature of the individual approaches. In the case of CE HRG, CE DQPM and SMASH we set it to $T_c = 158$ MeV since these approaches do not incooperate a cross-over.
	
	We first note that the results are in good agreement in each individual hadronic and partonic sector. The comparison between both Figures demonstrates the advantage of the $uds$- in contrast to the $BQS$-representation in the case of strongly-interacting matter because the discrepancies between the hadronic and partonic regime become apparent in the $uds$-basis. We observe that the diagonal light-quark coefficients grow in magnitude for decreasing temperatures on the hadronic side. Meanwhile, those from the partonic side approach zero. This growth in magnitude is also visible in the cross-conductivity $\kappa_{ud}$. Further, the cross-conductivities $\kappa_{us}$ and $\kappa_{ds}$ show non-trivial behavior consistent across he hadronic approaches, while on the partonic side they are all zero. This discrepancy is also evident in the comparison of the electric conductivity $\kappa_{QQ}$, which is also clear from Eq.\,\eqref{eq:relations_diff_coeff_QCD} because the dominating contributors to $\kappa_{QQ}$ are the diagonal quark-flavor coefficients and $-\kappa_{ud}$. This also explains why the partonic $\kappa_{BQ}$ does not compare well to the hadronic result.
	
	One of the reasons why the predictions from lattice QCD may not catch these differences in behavior at low temperatures seems to lie in the set of assumptions made in connection with Eq.\,\eqref{eq:lattice_approx}, where it was assumed that the off-diagonal quark-flavor current-current correlations are not taken into account. This is suggested by the very good comparison to the CE (DQPM) calculations, where a QGP was assumed. As a consequence, contributions like for example the up-anti-down correlations are discarded which - due to the formation of charged pions - would normally dominate below the critical temperature. It further seems that the content of light quarks depletes in the LQCD computations at low temperatures. Here, the structure in $\kappa_{us}$ and $\kappa_{ds}$ is mainly contributed to by kaons and the lightest hyperons.  This is supported by the good comparison between the results from the CE method, which only accounts for the $\Lambda$- and $\Sigma$-hyperons, while the full SMASH computation includes all known hadronic particles and resonances up to a mass of 2 GeV\cite{SMASH:2016zqf}.
	
	It may also be interesting to analyze the individual influence of each quark-flavor on the conductivities. To do so, we present the ratios $\kappa_{uu}/\kappa_{dd}$, which is indicative of the isospin symmetry, and $\kappa_{ss}/\kappa_{\ell\ell}$, which expresses the impact of the strange-quark in contrast to the light quarks. In the case of perfect isospin symmetry it is $\kappa_{uu} = \kappa_{dd}$, and thus their ratio would be unity. In the matter of the strange-quark influence, we defined $\kappa_{\ell\ell} = ( \kappa_{uu} + \kappa_{dd} )/2$, and thus $\kappa_{ss}/\kappa_{\ell\ell}$ would be unity if the strange quark contributes equally as the light quarks do. Figure \ref{fig:quark_influences} shows such a comparison for results from SMASH, CE HRG, CE DQPM, and LQCD. In the latter case, we only show the results published in Ref.\,\cite{Aarts:2014nba} because in that paper the authors provided individual conductivities $\kappa_{\ell\ell}$ and $\kappa_{ss}$ in a $N_f = 2+1$ simulation. That work therefore provides direct information on the strange-quark influence. We observe that $\kappa_{uu}/\kappa_{dd}$ is close to unity across the shown approaches with the exception of CE HRG. This difference is explained by the cross sections used in Refs.\,\cite{Greif:2017byw,Fotakis:2019nbq} which did not necessarily obey isospin symmetry. For $\kappa_{ss}/\kappa_{\ell\ell}$ we see that, as expected, the light quarks dominate over the strange quarks up to $T_c$. For higher temperatures ratio approaches unity and thus all quarks equally contribute. From the behavior of the CE (HRG) calculation at high temperatures we can further tell that the consideration of strange degrees of freedom is not sufficient in that model.
	
	\begin{widetext}
		\begin{center}
			\begin{figure*}[h!]
				\centering
				\includegraphics[width=0.8\textwidth]{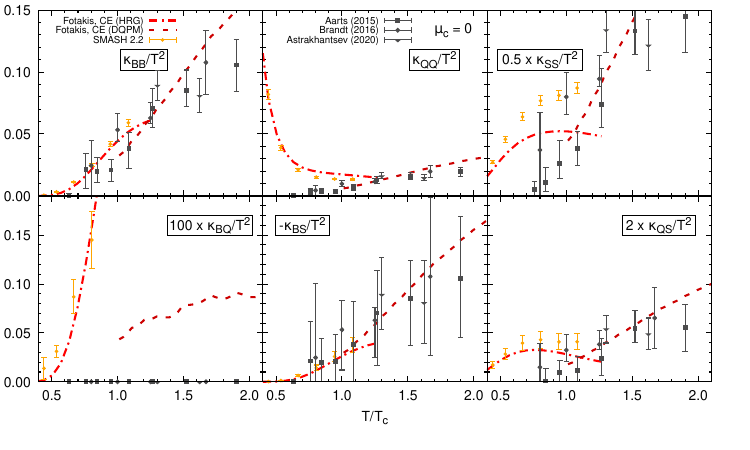}
				\vspace{-0.5cm}
				\caption{All six independent conductivities $\kappa_{qq'}/T^2$ (= $\sigma_{qq'}/T$) shown as function of $T/T_c$ in hadronic-charge ($BQS$) representation and at zero chemical potential, $\mu_c = 0$. Results are shown from hadronic and partonic approaches. For temperatures below $T_c = 158$ MeV we show results from SMASH \cite{SMASH:2016zqf} taken from Ref.\,\cite{Hammelmann:2023fqw} (orange rhomboids), and results from the Chapman-Enskog (CE) method for a $\pi KN\Lambda\Sigma$-gas \cite{Greif:2017byw,Fotakis:2019nbq} (HRG - red dash-dotted lines). For temperatures above $T_c = 158$ MeV we show results from the CE method for a quark-gluon gas \cite{Fotakis:2021diq} described by the DQPM \cite{Soloveva:2019xph} (DQPM - dark-red dashed lines). Further, the dark grey squares \cite{Aarts:2014nba}, circles \cite{Brandt:2015aqk}, and triangles \cite{Astrakhantsev:2019zkr} represent results evaluated according to Eqs.\,\eqref{eq:off_diag_kappa_to_diag_isopin_symm} from available calculations from lattice QCD under the assumption of equal influences of up-, down-, and strange-quarks. For the LQCD results we do not show errorbars for the off-diagonal coefficients which according to Eqs.\,\eqref{eq:lattice_QCD_Nf2+1} are identical to zero.}
				\label{fig:kappa_BQS_Nf2+1_BQS}
			\end{figure*}
			\begin{figure*}[h!]
				\centering
				\includegraphics[width=0.8\textwidth]{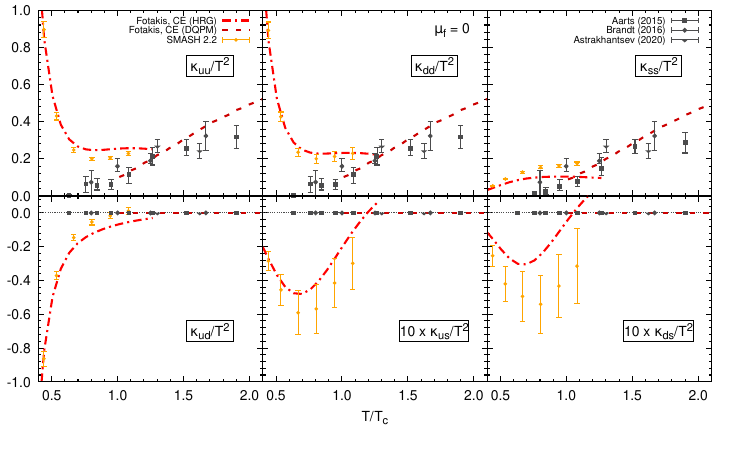}
				\vspace{-0.5cm}
				\caption{Same as in Figure \ref{fig:kappa_BQS_Nf2+1_BQS} but the conductivities are in quark-flavor ($uds$) representation.}
				\label{fig:kappa_uds_Nf2+1_uds}
			\end{figure*}
			\begin{figure*}[h!]
				\centering
				\includegraphics[width=0.7\textwidth]{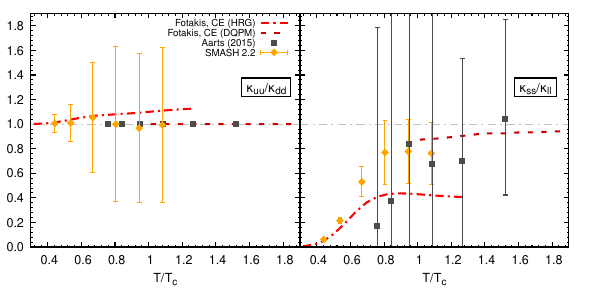}
				\vspace{-0.5cm}
				\caption{Ratios of various quark-flavor conductivities as function of $T/T_c$ at vanishing chemical potential. Left: ratio of up-quark conductivity over down-quark conductivity. Right: ratio of strange-quark conductivity over light-quark conductivity. We show results from SMASH \cite{Hammelmann:2023fqw} (orange rhomboids), the CE method for a hadron gas \cite{Greif:2017byw,Fotakis:2019nbq} (CE HRG - red dash-dotted line) and a partonic system \cite{Fotakis:2021diq} (CE DQPM - dark red dashed line). Further, results from LQCD are shown \cite{Aarts:2014nba} (dark grey squares). The grey dash-dotted line marks unity.}
				\label{fig:quark_influences}
			\end{figure*}
		\end{center}
	\end{widetext}

	\quad\newline\quad\newline\quad\newline
	
	\section{Conclusion}
	\label{sec:Conclusion}
	
	In this paper, we reviewed the diffusion or conductivity matrix of strongly-interacting matter, and its transformation with respect to a change of basis in charge representation space. From this we achieved exact relations between the cross-conductivities and the diagonal conductivities, with the help of which we could discuss these coefficients from available lattice QCD results. 
	
	We conjectured that the assumption of neglecting the off-diagonal contributions to the quark-quark current correlator in the LQCD analysis may lead to unexpected deviations to hadronic models at lower temperatures, and we would like to see these assumptions relaxed if possible. Apart from that, we found that the available calculations of the full conductivity matrix are in good agreement with each other.
	
	We gave the first demonstration that the anaylsis of the conductivities in the quark-flavor representation provides a novel tool to analyze the influences of the quarks on the transport properties of the medium and how they may be distributed onto hadronic bound states. We have indications of non-trivial structures in the off-diagonal coefficients close to $T_c$ from the hadronic approaches, which are so far not covered by the available LQCD analysis. We argue that especially the off-diagonal coefficients in quark-flavor representation may deliver some insight into the chemical composition of the strongly-interacting medium. It would be very interesting to have results from partonic models, which also incorporate a crossover or a phase transition, in order to investigate how these non-trivial structures - visible in the hadronic analysis - arise on the partonic side. From the observations above, we conjecture that the (off-diagonal) coefficients may be sensitive to the presence of a critical endpoint. Thus an analysis of lattice calculations at finite chemical potential could provide valuable insights. Finally, recent work by German LQCD groups \cite{Bala:2023iqu} indicate that hadronic-like excitations may be present at very high temperatures, where traditionally one would suspect a pure, deconfined partonic phase. Contrary to a QGP such excitations in the medium would lead to non-zero off-diagonal conductivities.
	
	\section*{Acknowledgments}
	
	The authors thank O.\,Philipsen and O.\,Kaczmarek for fruitful discussions. They acknowledge support by the 
	Deutsche Forschungsgemeinschaft (DFG, German Research Foundation) through the CRC-TR 211 ``Strong-interaction matter under extreme conditions'' -- project number 
	315477589 -- TRR 211. J.A.F.\ acknowledges support from the Helmholtz Graduate School for Heavy-Ion Research.

	\bibliographystyle{apsrev4-1}
	\bibliography{paper.bib} 

\begin{thebibliography}{34}%
\makeatletter
\providecommand \@ifxundefined [1]{%
 \@ifx{#1\undefined}
}%
\providecommand \@ifnum [1]{%
 \ifnum #1\expandafter \@firstoftwo
 \else \expandafter \@secondoftwo
 \fi
}%
\providecommand \@ifx [1]{%
 \ifx #1\expandafter \@firstoftwo
 \else \expandafter \@secondoftwo
 \fi
}%
\providecommand \natexlab [1]{#1}%
\providecommand \enquote  [1]{``#1''}%
\providecommand \bibnamefont  [1]{#1}%
\providecommand \bibfnamefont [1]{#1}%
\providecommand \citenamefont [1]{#1}%
\providecommand \href@noop [0]{\@secondoftwo}%
\providecommand \href [0]{\begingroup \@sanitize@url \@href}%
\providecommand \@href[1]{\@@startlink{#1}\@@href}%
\providecommand \@@href[1]{\endgroup#1\@@endlink}%
\providecommand \@sanitize@url [0]{\catcode `\\12\catcode `\$12\catcode
  `\&12\catcode `\#12\catcode `\^12\catcode `\_12\catcode `\%12\relax}%
\providecommand \@@startlink[1]{}%
\providecommand \@@endlink[0]{}%
\providecommand \url  [0]{\begingroup\@sanitize@url \@url }%
\providecommand \@url [1]{\endgroup\@href {#1}{\urlprefix }}%
\providecommand \urlprefix  [0]{URL }%
\providecommand \Eprint [0]{\href }%
\providecommand \doibase [0]{http://dx.doi.org/}%
\providecommand \selectlanguage [0]{\@gobble}%
\providecommand \bibinfo  [0]{\@secondoftwo}%
\providecommand \bibfield  [0]{\@secondoftwo}%
\providecommand \translation [1]{[#1]}%
\providecommand \BibitemOpen [0]{}%
\providecommand \bibitemStop [0]{}%
\providecommand \bibitemNoStop [0]{.\EOS\space}%
\providecommand \EOS [0]{\spacefactor3000\relax}%
\providecommand \BibitemShut  [1]{\csname bibitem#1\endcsname}%
\let\auto@bib@innerbib\@empty
\bibitem [{\citenamefont {Puglisi}\ \emph {et~al.}(2015)\citenamefont
  {Puglisi}, \citenamefont {Plumari},\ and\ \citenamefont
  {Greco}}]{Puglisi:2014pda}%
  \BibitemOpen
  \bibfield  {author} {\bibinfo {author} {\bibfnamefont {A.}~\bibnamefont
  {Puglisi}}, \bibinfo {author} {\bibfnamefont {S.}~\bibnamefont {Plumari}}, \
  and\ \bibinfo {author} {\bibfnamefont {V.}~\bibnamefont {Greco}},\ }\href
  {\doibase 10.1016/j.physletb.2015.10.070} {\bibfield  {journal} {\bibinfo
  {journal} {Phys. Lett. B}\ }\textbf {\bibinfo {volume} {751}},\ \bibinfo
  {pages} {326} (\bibinfo {year} {2015})},\ \Eprint
  {http://arxiv.org/abs/1407.2559} {arXiv:1407.2559 [hep-ph]} \BibitemShut
  {NoStop}%
\bibitem [{\citenamefont {Puglisi}\ \emph {et~al.}(2014)\citenamefont
  {Puglisi}, \citenamefont {Plumari},\ and\ \citenamefont
  {Greco}}]{Puglisi:2014sha}%
  \BibitemOpen
  \bibfield  {author} {\bibinfo {author} {\bibfnamefont {A.}~\bibnamefont
  {Puglisi}}, \bibinfo {author} {\bibfnamefont {S.}~\bibnamefont {Plumari}}, \
  and\ \bibinfo {author} {\bibfnamefont {V.}~\bibnamefont {Greco}},\ }\href
  {\doibase 10.1103/PhysRevD.90.114009} {\bibfield  {journal} {\bibinfo
  {journal} {Phys. Rev. D}\ }\textbf {\bibinfo {volume} {90}},\ \bibinfo
  {pages} {114009} (\bibinfo {year} {2014})},\ \Eprint
  {http://arxiv.org/abs/1408.7043} {arXiv:1408.7043 [hep-ph]} \BibitemShut
  {NoStop}%
\bibitem [{\citenamefont {Thakur}\ \emph {et~al.}(2017)\citenamefont {Thakur},
  \citenamefont {Srivastava}, \citenamefont {Kadam}, \citenamefont {George},\
  and\ \citenamefont {Mishra}}]{Thakur:2017hfc}%
  \BibitemOpen
  \bibfield  {author} {\bibinfo {author} {\bibfnamefont {L.}~\bibnamefont
  {Thakur}}, \bibinfo {author} {\bibfnamefont {P.~K.}\ \bibnamefont
  {Srivastava}}, \bibinfo {author} {\bibfnamefont {G.~P.}\ \bibnamefont
  {Kadam}}, \bibinfo {author} {\bibfnamefont {M.}~\bibnamefont {George}}, \
  and\ \bibinfo {author} {\bibfnamefont {H.}~\bibnamefont {Mishra}},\ }\href
  {\doibase 10.1103/PhysRevD.95.096009} {\bibfield  {journal} {\bibinfo
  {journal} {Phys. Rev. D}\ }\textbf {\bibinfo {volume} {95}},\ \bibinfo
  {pages} {096009} (\bibinfo {year} {2017})},\ \Eprint
  {http://arxiv.org/abs/1703.03142} {arXiv:1703.03142 [hep-ph]} \BibitemShut
  {NoStop}%
\bibitem [{\citenamefont {Greif}\ \emph {et~al.}(2016)\citenamefont {Greif},
  \citenamefont {Greiner},\ and\ \citenamefont {Denicol}}]{Greif:2016skc}%
  \BibitemOpen
  \bibfield  {author} {\bibinfo {author} {\bibfnamefont {M.}~\bibnamefont
  {Greif}}, \bibinfo {author} {\bibfnamefont {C.}~\bibnamefont {Greiner}}, \
  and\ \bibinfo {author} {\bibfnamefont {G.~S.}\ \bibnamefont {Denicol}},\
  }\href {\doibase 10.1103/PhysRevD.93.096012} {\bibfield  {journal} {\bibinfo
  {journal} {Phys. Rev. D}\ }\textbf {\bibinfo {volume} {93}},\ \bibinfo
  {pages} {096012} (\bibinfo {year} {2016})},\ \bibinfo {note} {[Erratum:
  Phys.Rev.D 96, 059902 (2017)]},\ \Eprint {http://arxiv.org/abs/1602.05085}
  {arXiv:1602.05085 [nucl-th]} \BibitemShut {NoStop}%
\bibitem [{\citenamefont {Greif}\ \emph {et~al.}(2018)\citenamefont {Greif},
  \citenamefont {Fotakis}, \citenamefont {Denicol},\ and\ \citenamefont
  {Greiner}}]{Greif:2017byw}%
  \BibitemOpen
  \bibfield  {author} {\bibinfo {author} {\bibfnamefont {M.}~\bibnamefont
  {Greif}}, \bibinfo {author} {\bibfnamefont {J.~A.}\ \bibnamefont {Fotakis}},
  \bibinfo {author} {\bibfnamefont {G.~S.}\ \bibnamefont {Denicol}}, \ and\
  \bibinfo {author} {\bibfnamefont {C.}~\bibnamefont {Greiner}},\ }\href
  {\doibase 10.1103/PhysRevLett.120.242301} {\bibfield  {journal} {\bibinfo
  {journal} {Phys. Rev. Lett.}\ }\textbf {\bibinfo {volume} {120}},\ \bibinfo
  {pages} {242301} (\bibinfo {year} {2018})},\ \Eprint
  {http://arxiv.org/abs/1711.08680} {arXiv:1711.08680 [hep-ph]} \BibitemShut
  {NoStop}%
\bibitem [{\citenamefont {Fotakis}\ \emph {et~al.}(2020)\citenamefont
  {Fotakis}, \citenamefont {Greif}, \citenamefont {Greiner}, \citenamefont
  {Denicol},\ and\ \citenamefont {Niemi}}]{Fotakis:2019nbq}%
  \BibitemOpen
  \bibfield  {author} {\bibinfo {author} {\bibfnamefont {J.~A.}\ \bibnamefont
  {Fotakis}}, \bibinfo {author} {\bibfnamefont {M.}~\bibnamefont {Greif}},
  \bibinfo {author} {\bibfnamefont {C.}~\bibnamefont {Greiner}}, \bibinfo
  {author} {\bibfnamefont {G.~S.}\ \bibnamefont {Denicol}}, \ and\ \bibinfo
  {author} {\bibfnamefont {H.}~\bibnamefont {Niemi}},\ }\href {\doibase
  10.1103/PhysRevD.101.076007} {\bibfield  {journal} {\bibinfo  {journal}
  {Phys. Rev. D}\ }\textbf {\bibinfo {volume} {101}},\ \bibinfo {pages}
  {076007} (\bibinfo {year} {2020})},\ \Eprint
  {http://arxiv.org/abs/1912.09103} {arXiv:1912.09103 [hep-ph]} \BibitemShut
  {NoStop}%
\bibitem [{\citenamefont {Fotakis}\ \emph {et~al.}(2021)\citenamefont
  {Fotakis}, \citenamefont {Soloveva}, \citenamefont {Greiner}, \citenamefont
  {Kaczmarek},\ and\ \citenamefont {Bratkovskaya}}]{Fotakis:2021diq}%
  \BibitemOpen
  \bibfield  {author} {\bibinfo {author} {\bibfnamefont {J.~A.}\ \bibnamefont
  {Fotakis}}, \bibinfo {author} {\bibfnamefont {O.}~\bibnamefont {Soloveva}},
  \bibinfo {author} {\bibfnamefont {C.}~\bibnamefont {Greiner}}, \bibinfo
  {author} {\bibfnamefont {O.}~\bibnamefont {Kaczmarek}}, \ and\ \bibinfo
  {author} {\bibfnamefont {E.}~\bibnamefont {Bratkovskaya}},\ }\href {\doibase
  10.1103/PhysRevD.104.034014} {\bibfield  {journal} {\bibinfo  {journal}
  {Phys. Rev. D}\ }\textbf {\bibinfo {volume} {104}},\ \bibinfo {pages}
  {034014} (\bibinfo {year} {2021})},\ \Eprint
  {http://arxiv.org/abs/2102.08140} {arXiv:2102.08140 [hep-ph]} \BibitemShut
  {NoStop}%
\bibitem [{\citenamefont {Greif}\ \emph {et~al.}(2014)\citenamefont {Greif},
  \citenamefont {Bouras}, \citenamefont {Greiner},\ and\ \citenamefont
  {Xu}}]{Greif:2014oia}%
  \BibitemOpen
  \bibfield  {author} {\bibinfo {author} {\bibfnamefont {M.}~\bibnamefont
  {Greif}}, \bibinfo {author} {\bibfnamefont {I.}~\bibnamefont {Bouras}},
  \bibinfo {author} {\bibfnamefont {C.}~\bibnamefont {Greiner}}, \ and\
  \bibinfo {author} {\bibfnamefont {Z.}~\bibnamefont {Xu}},\ }\href {\doibase
  10.1103/PhysRevD.90.094014} {\bibfield  {journal} {\bibinfo  {journal} {Phys.
  Rev. D}\ }\textbf {\bibinfo {volume} {90}},\ \bibinfo {pages} {094014}
  (\bibinfo {year} {2014})},\ \Eprint {http://arxiv.org/abs/1408.7049}
  {arXiv:1408.7049 [nucl-th]} \BibitemShut {NoStop}%
\bibitem [{\citenamefont {Hammelmann}\ \emph {et~al.}(2019)\citenamefont
  {Hammelmann}, \citenamefont {Torres-Rincon}, \citenamefont {Rose},
  \citenamefont {Greif},\ and\ \citenamefont {Elfner}}]{Hammelmann:2018ath}%
  \BibitemOpen
  \bibfield  {author} {\bibinfo {author} {\bibfnamefont {J.}~\bibnamefont
  {Hammelmann}}, \bibinfo {author} {\bibfnamefont {J.~M.}\ \bibnamefont
  {Torres-Rincon}}, \bibinfo {author} {\bibfnamefont {J.-B.}\ \bibnamefont
  {Rose}}, \bibinfo {author} {\bibfnamefont {M.}~\bibnamefont {Greif}}, \ and\
  \bibinfo {author} {\bibfnamefont {H.}~\bibnamefont {Elfner}},\ }\href
  {\doibase 10.1103/PhysRevD.99.076015} {\bibfield  {journal} {\bibinfo
  {journal} {Phys. Rev. D}\ }\textbf {\bibinfo {volume} {99}},\ \bibinfo
  {pages} {076015} (\bibinfo {year} {2019})},\ \Eprint
  {http://arxiv.org/abs/1810.12527} {arXiv:1810.12527 [hep-ph]} \BibitemShut
  {NoStop}%
\bibitem [{\citenamefont {Rose}\ \emph {et~al.}(2020)\citenamefont {Rose},
  \citenamefont {Greif}, \citenamefont {Hammelmann}, \citenamefont {Fotakis},
  \citenamefont {Denicol}, \citenamefont {Elfner},\ and\ \citenamefont
  {Greiner}}]{Rose:2020sjv}%
  \BibitemOpen
  \bibfield  {author} {\bibinfo {author} {\bibfnamefont {J.-B.}\ \bibnamefont
  {Rose}}, \bibinfo {author} {\bibfnamefont {M.}~\bibnamefont {Greif}},
  \bibinfo {author} {\bibfnamefont {J.}~\bibnamefont {Hammelmann}}, \bibinfo
  {author} {\bibfnamefont {J.~A.}\ \bibnamefont {Fotakis}}, \bibinfo {author}
  {\bibfnamefont {G.~S.}\ \bibnamefont {Denicol}}, \bibinfo {author}
  {\bibfnamefont {H.}~\bibnamefont {Elfner}}, \ and\ \bibinfo {author}
  {\bibfnamefont {C.}~\bibnamefont {Greiner}},\ }\href {\doibase
  10.1103/PhysRevD.101.114028} {\bibfield  {journal} {\bibinfo  {journal}
  {Phys. Rev. D}\ }\textbf {\bibinfo {volume} {101}},\ \bibinfo {pages}
  {114028} (\bibinfo {year} {2020})},\ \Eprint
  {http://arxiv.org/abs/2001.10606} {arXiv:2001.10606 [nucl-th]} \BibitemShut
  {NoStop}%
\bibitem [{\citenamefont {Arnold}\ \emph {et~al.}(2000)\citenamefont {Arnold},
  \citenamefont {Moore},\ and\ \citenamefont {Yaffe}}]{Arnold:2000dr}%
  \BibitemOpen
  \bibfield  {author} {\bibinfo {author} {\bibfnamefont {P.~B.}\ \bibnamefont
  {Arnold}}, \bibinfo {author} {\bibfnamefont {G.~D.}\ \bibnamefont {Moore}}, \
  and\ \bibinfo {author} {\bibfnamefont {L.~G.}\ \bibnamefont {Yaffe}},\ }\href
  {\doibase 10.1088/1126-6708/2000/11/001} {\bibfield  {journal} {\bibinfo
  {journal} {JHEP}\ }\textbf {\bibinfo {volume} {11}},\ \bibinfo {pages} {001}
  (\bibinfo {year} {2000})},\ \Eprint {http://arxiv.org/abs/hep-ph/0010177}
  {arXiv:hep-ph/0010177} \BibitemShut {NoStop}%
\bibitem [{\citenamefont {Arnold}\ \emph {et~al.}(2003)\citenamefont {Arnold},
  \citenamefont {Moore},\ and\ \citenamefont {Yaffe}}]{Arnold:2003zc}%
  \BibitemOpen
  \bibfield  {author} {\bibinfo {author} {\bibfnamefont {P.~B.}\ \bibnamefont
  {Arnold}}, \bibinfo {author} {\bibfnamefont {G.~D.}\ \bibnamefont {Moore}}, \
  and\ \bibinfo {author} {\bibfnamefont {L.~G.}\ \bibnamefont {Yaffe}},\ }\href
  {\doibase 10.1088/1126-6708/2003/05/051} {\bibfield  {journal} {\bibinfo
  {journal} {JHEP}\ }\textbf {\bibinfo {volume} {05}},\ \bibinfo {pages} {051}
  (\bibinfo {year} {2003})},\ \Eprint {http://arxiv.org/abs/hep-ph/0302165}
  {arXiv:hep-ph/0302165} \BibitemShut {NoStop}%
\bibitem [{\citenamefont {Torres-Rincon}(2012)}]{Torres-Rincon:2012sda}%
  \BibitemOpen
  \bibfield  {author} {\bibinfo {author} {\bibfnamefont {J.~M.}\ \bibnamefont
  {Torres-Rincon}},\ }\emph {\bibinfo {title} {{Hadronic transport coefficients
  from effective field theories}}},\ \href {\doibase 10.1007/978-3-319-00425-9}
  {Ph.D. thesis},\ \bibinfo  {school} {UCM, Madrid, Dept. Phys., UCM,
  Somosaguas} (\bibinfo {year} {2012}),\ \Eprint
  {http://arxiv.org/abs/1205.0782} {arXiv:1205.0782 [hep-ph]} \BibitemShut
  {NoStop}%
\bibitem [{\citenamefont {Soloveva}\ \emph {et~al.}(2020)\citenamefont
  {Soloveva}, \citenamefont {Moreau},\ and\ \citenamefont
  {Bratkovskaya}}]{Soloveva:2019xph}%
  \BibitemOpen
  \bibfield  {author} {\bibinfo {author} {\bibfnamefont {O.}~\bibnamefont
  {Soloveva}}, \bibinfo {author} {\bibfnamefont {P.}~\bibnamefont {Moreau}}, \
  and\ \bibinfo {author} {\bibfnamefont {E.}~\bibnamefont {Bratkovskaya}},\
  }\href {\doibase 10.1103/PhysRevC.101.045203} {\bibfield  {journal} {\bibinfo
   {journal} {Phys. Rev. C}\ }\textbf {\bibinfo {volume} {101}},\ \bibinfo
  {pages} {045203} (\bibinfo {year} {2020})},\ \Eprint
  {http://arxiv.org/abs/1911.08547} {arXiv:1911.08547 [nucl-th]} \BibitemShut
  {NoStop}%
\bibitem [{\citenamefont {Finazzo}\ and\ \citenamefont
  {Noronha}(2014)}]{Finazzo:2013efa}%
  \BibitemOpen
  \bibfield  {author} {\bibinfo {author} {\bibfnamefont {S.~I.}\ \bibnamefont
  {Finazzo}}\ and\ \bibinfo {author} {\bibfnamefont {J.}~\bibnamefont
  {Noronha}},\ }\href {\doibase 10.1103/PhysRevD.89.106008} {\bibfield
  {journal} {\bibinfo  {journal} {Phys. Rev. D}\ }\textbf {\bibinfo {volume}
  {89}},\ \bibinfo {pages} {106008} (\bibinfo {year} {2014})},\ \Eprint
  {http://arxiv.org/abs/1311.6675} {arXiv:1311.6675 [hep-th]} \BibitemShut
  {NoStop}%
\bibitem [{\citenamefont {Rougemont}\ \emph {et~al.}(2015)\citenamefont
  {Rougemont}, \citenamefont {Noronha},\ and\ \citenamefont
  {Noronha-Hostler}}]{Rougemont:2015ona}%
  \BibitemOpen
  \bibfield  {author} {\bibinfo {author} {\bibfnamefont {R.}~\bibnamefont
  {Rougemont}}, \bibinfo {author} {\bibfnamefont {J.}~\bibnamefont {Noronha}},
  \ and\ \bibinfo {author} {\bibfnamefont {J.}~\bibnamefont
  {Noronha-Hostler}},\ }\href {\doibase 10.1103/PhysRevLett.115.202301}
  {\bibfield  {journal} {\bibinfo  {journal} {Phys. Rev. Lett.}\ }\textbf
  {\bibinfo {volume} {115}},\ \bibinfo {pages} {202301} (\bibinfo {year}
  {2015})},\ \Eprint {http://arxiv.org/abs/1507.06972} {arXiv:1507.06972
  [hep-ph]} \BibitemShut {NoStop}%
\bibitem [{\citenamefont {Rougemont}\ \emph {et~al.}(2017)\citenamefont
  {Rougemont}, \citenamefont {Critelli}, \citenamefont {Noronha-Hostler},
  \citenamefont {Noronha},\ and\ \citenamefont {Ratti}}]{Rougemont:2017tlu}%
  \BibitemOpen
  \bibfield  {author} {\bibinfo {author} {\bibfnamefont {R.}~\bibnamefont
  {Rougemont}}, \bibinfo {author} {\bibfnamefont {R.}~\bibnamefont {Critelli}},
  \bibinfo {author} {\bibfnamefont {J.}~\bibnamefont {Noronha-Hostler}},
  \bibinfo {author} {\bibfnamefont {J.}~\bibnamefont {Noronha}}, \ and\
  \bibinfo {author} {\bibfnamefont {C.}~\bibnamefont {Ratti}},\ }\href
  {\doibase 10.1103/PhysRevD.96.014032} {\bibfield  {journal} {\bibinfo
  {journal} {Phys. Rev. D}\ }\textbf {\bibinfo {volume} {96}},\ \bibinfo
  {pages} {014032} (\bibinfo {year} {2017})},\ \Eprint
  {http://arxiv.org/abs/1704.05558} {arXiv:1704.05558 [hep-ph]} \BibitemShut
  {NoStop}%
\bibitem [{\citenamefont {Aarts}\ \emph {et~al.}(2007)\citenamefont {Aarts},
  \citenamefont {Allton}, \citenamefont {Foley}, \citenamefont {Hands},\ and\
  \citenamefont {Kim}}]{Aarts:2007wj}%
  \BibitemOpen
  \bibfield  {author} {\bibinfo {author} {\bibfnamefont {G.}~\bibnamefont
  {Aarts}}, \bibinfo {author} {\bibfnamefont {C.}~\bibnamefont {Allton}},
  \bibinfo {author} {\bibfnamefont {J.}~\bibnamefont {Foley}}, \bibinfo
  {author} {\bibfnamefont {S.}~\bibnamefont {Hands}}, \ and\ \bibinfo {author}
  {\bibfnamefont {S.}~\bibnamefont {Kim}},\ }\href {\doibase
  10.1103/PhysRevLett.99.022002} {\bibfield  {journal} {\bibinfo  {journal}
  {Phys. Rev. Lett.}\ }\textbf {\bibinfo {volume} {99}},\ \bibinfo {pages}
  {022002} (\bibinfo {year} {2007})},\ \Eprint
  {http://arxiv.org/abs/hep-lat/0703008} {arXiv:hep-lat/0703008} \BibitemShut
  {NoStop}%
\bibitem [{\citenamefont {Brandt}\ \emph {et~al.}(2013)\citenamefont {Brandt},
  \citenamefont {Francis}, \citenamefont {Meyer},\ and\ \citenamefont
  {Wittig}}]{Brandt:2012jc}%
  \BibitemOpen
  \bibfield  {author} {\bibinfo {author} {\bibfnamefont {B.~B.}\ \bibnamefont
  {Brandt}}, \bibinfo {author} {\bibfnamefont {A.}~\bibnamefont {Francis}},
  \bibinfo {author} {\bibfnamefont {H.~B.}\ \bibnamefont {Meyer}}, \ and\
  \bibinfo {author} {\bibfnamefont {H.}~\bibnamefont {Wittig}},\ }\href
  {\doibase 10.1007/JHEP03(2013)100} {\bibfield  {journal} {\bibinfo  {journal}
  {JHEP}\ }\textbf {\bibinfo {volume} {03}},\ \bibinfo {pages} {100} (\bibinfo
  {year} {2013})},\ \Eprint {http://arxiv.org/abs/1212.4200} {arXiv:1212.4200
  [hep-lat]} \BibitemShut {NoStop}%
\bibitem [{\citenamefont {Amato}\ \emph {et~al.}(2013)\citenamefont {Amato},
  \citenamefont {Aarts}, \citenamefont {Allton}, \citenamefont {Giudice},
  \citenamefont {Hands},\ and\ \citenamefont {Skullerud}}]{Amato:2013naa}%
  \BibitemOpen
  \bibfield  {author} {\bibinfo {author} {\bibfnamefont {A.}~\bibnamefont
  {Amato}}, \bibinfo {author} {\bibfnamefont {G.}~\bibnamefont {Aarts}},
  \bibinfo {author} {\bibfnamefont {C.}~\bibnamefont {Allton}}, \bibinfo
  {author} {\bibfnamefont {P.}~\bibnamefont {Giudice}}, \bibinfo {author}
  {\bibfnamefont {S.}~\bibnamefont {Hands}}, \ and\ \bibinfo {author}
  {\bibfnamefont {J.-I.}\ \bibnamefont {Skullerud}},\ }\href {\doibase
  10.1103/PhysRevLett.111.172001} {\bibfield  {journal} {\bibinfo  {journal}
  {Phys. Rev. Lett.}\ }\textbf {\bibinfo {volume} {111}},\ \bibinfo {pages}
  {172001} (\bibinfo {year} {2013})},\ \Eprint {http://arxiv.org/abs/1307.6763}
  {arXiv:1307.6763 [hep-lat]} \BibitemShut {NoStop}%
\bibitem [{\citenamefont {Aarts}\ \emph {et~al.}(2015)\citenamefont {Aarts},
  \citenamefont {Allton}, \citenamefont {Amato}, \citenamefont {Giudice},
  \citenamefont {Hands},\ and\ \citenamefont {Skullerud}}]{Aarts:2014nba}%
  \BibitemOpen
  \bibfield  {author} {\bibinfo {author} {\bibfnamefont {G.}~\bibnamefont
  {Aarts}}, \bibinfo {author} {\bibfnamefont {C.}~\bibnamefont {Allton}},
  \bibinfo {author} {\bibfnamefont {A.}~\bibnamefont {Amato}}, \bibinfo
  {author} {\bibfnamefont {P.}~\bibnamefont {Giudice}}, \bibinfo {author}
  {\bibfnamefont {S.}~\bibnamefont {Hands}}, \ and\ \bibinfo {author}
  {\bibfnamefont {J.-I.}\ \bibnamefont {Skullerud}},\ }\href {\doibase
  10.1007/JHEP02(2015)186} {\bibfield  {journal} {\bibinfo  {journal} {JHEP}\
  }\textbf {\bibinfo {volume} {02}},\ \bibinfo {pages} {186} (\bibinfo {year}
  {2015})},\ \Eprint {http://arxiv.org/abs/1412.6411} {arXiv:1412.6411
  [hep-lat]} \BibitemShut {NoStop}%
\bibitem [{\citenamefont {Brandt}\ \emph {et~al.}(2016)\citenamefont {Brandt},
  \citenamefont {Francis}, \citenamefont {J\"ager},\ and\ \citenamefont
  {Meyer}}]{Brandt:2015aqk}%
  \BibitemOpen
  \bibfield  {author} {\bibinfo {author} {\bibfnamefont {B.~B.}\ \bibnamefont
  {Brandt}}, \bibinfo {author} {\bibfnamefont {A.}~\bibnamefont {Francis}},
  \bibinfo {author} {\bibfnamefont {B.}~\bibnamefont {J\"ager}}, \ and\
  \bibinfo {author} {\bibfnamefont {H.~B.}\ \bibnamefont {Meyer}},\ }\href
  {\doibase 10.1103/PhysRevD.93.054510} {\bibfield  {journal} {\bibinfo
  {journal} {Phys. Rev. D}\ }\textbf {\bibinfo {volume} {93}},\ \bibinfo
  {pages} {054510} (\bibinfo {year} {2016})},\ \Eprint
  {http://arxiv.org/abs/1512.07249} {arXiv:1512.07249 [hep-lat]} \BibitemShut
  {NoStop}%
\bibitem [{\citenamefont {Ding}\ \emph {et~al.}(2016)\citenamefont {Ding},
  \citenamefont {Kaczmarek},\ and\ \citenamefont {Meyer}}]{Ding:2016hua}%
  \BibitemOpen
  \bibfield  {author} {\bibinfo {author} {\bibfnamefont {H.-T.}\ \bibnamefont
  {Ding}}, \bibinfo {author} {\bibfnamefont {O.}~\bibnamefont {Kaczmarek}}, \
  and\ \bibinfo {author} {\bibfnamefont {F.}~\bibnamefont {Meyer}},\ }\href
  {\doibase 10.1103/PhysRevD.94.034504} {\bibfield  {journal} {\bibinfo
  {journal} {Phys. Rev. D}\ }\textbf {\bibinfo {volume} {94}},\ \bibinfo
  {pages} {034504} (\bibinfo {year} {2016})},\ \Eprint
  {http://arxiv.org/abs/1604.06712} {arXiv:1604.06712 [hep-lat]} \BibitemShut
  {NoStop}%
\bibitem [{\citenamefont {Astrakhantsev}\ \emph {et~al.}(2020)\citenamefont
  {Astrakhantsev}, \citenamefont {Braguta}, \citenamefont {D'Elia},
  \citenamefont {Kotov}, \citenamefont {Nikolaev},\ and\ \citenamefont
  {Sanfilippo}}]{Astrakhantsev:2019zkr}%
  \BibitemOpen
  \bibfield  {author} {\bibinfo {author} {\bibfnamefont {N.}~\bibnamefont
  {Astrakhantsev}}, \bibinfo {author} {\bibfnamefont {V.~V.}\ \bibnamefont
  {Braguta}}, \bibinfo {author} {\bibfnamefont {M.}~\bibnamefont {D'Elia}},
  \bibinfo {author} {\bibfnamefont {A.~Y.}\ \bibnamefont {Kotov}}, \bibinfo
  {author} {\bibfnamefont {A.~A.}\ \bibnamefont {Nikolaev}}, \ and\ \bibinfo
  {author} {\bibfnamefont {F.}~\bibnamefont {Sanfilippo}},\ }\href {\doibase
  10.1103/PhysRevD.102.054516} {\bibfield  {journal} {\bibinfo  {journal}
  {Phys. Rev. D}\ }\textbf {\bibinfo {volume} {102}},\ \bibinfo {pages}
  {054516} (\bibinfo {year} {2020})},\ \Eprint
  {http://arxiv.org/abs/1910.08516} {arXiv:1910.08516 [hep-lat]} \BibitemShut
  {NoStop}%
\bibitem [{\citenamefont {Hammelmann}\ \emph {et~al.}(2023)\citenamefont
  {Hammelmann}, \citenamefont {Staudenmaier},\ and\ \citenamefont
  {Elfner}}]{Hammelmann:2023fqw}%
  \BibitemOpen
  \bibfield  {author} {\bibinfo {author} {\bibfnamefont {J.}~\bibnamefont
  {Hammelmann}}, \bibinfo {author} {\bibfnamefont {J.}~\bibnamefont
  {Staudenmaier}}, \ and\ \bibinfo {author} {\bibfnamefont {H.}~\bibnamefont
  {Elfner}},\ }\href@noop {} {\  (\bibinfo {year} {2023})},\ \Eprint
  {http://arxiv.org/abs/2307.15606} {arXiv:2307.15606 [nucl-th]} \BibitemShut
  {NoStop}%
\bibitem [{\citenamefont {De~Groot}(1980)}]{DeGroot:1980dk}%
  \BibitemOpen
  \bibfield  {author} {\bibinfo {author} {\bibfnamefont {S.~R.}\ \bibnamefont
  {De~Groot}},\ }\href@noop {} {\emph {\bibinfo {title} {{Relativistic Kinetic
  Theory. Principles and Applications}}}},\ edited by\ \bibinfo {editor}
  {\bibfnamefont {W.~A.}\ \bibnamefont {Van~Leeuwen}}\ and\ \bibinfo {editor}
  {\bibfnamefont {C.~G.}\ \bibnamefont {Van~Weert}}\ (\bibinfo {year}
  {1980})\BibitemShut {NoStop}%
\bibitem [{\citenamefont {Monnai}\ and\ \citenamefont
  {Hirano}(2010)}]{Monnai:2010qp}%
  \BibitemOpen
  \bibfield  {author} {\bibinfo {author} {\bibfnamefont {A.}~\bibnamefont
  {Monnai}}\ and\ \bibinfo {author} {\bibfnamefont {T.}~\bibnamefont
  {Hirano}},\ }\href {\doibase 10.1016/j.nuclphysa.2010.08.002} {\bibfield
  {journal} {\bibinfo  {journal} {Nucl. Phys. A}\ }\textbf {\bibinfo {volume}
  {847}},\ \bibinfo {pages} {283} (\bibinfo {year} {2010})},\ \Eprint
  {http://arxiv.org/abs/1003.3087} {arXiv:1003.3087 [nucl-th]} \BibitemShut
  {NoStop}%
\bibitem [{\citenamefont {Kikuchi}\ \emph {et~al.}(2015)\citenamefont
  {Kikuchi}, \citenamefont {Tsumura},\ and\ \citenamefont
  {Kunihiro}}]{Kikuchi:2015swa}%
  \BibitemOpen
  \bibfield  {author} {\bibinfo {author} {\bibfnamefont {Y.}~\bibnamefont
  {Kikuchi}}, \bibinfo {author} {\bibfnamefont {K.}~\bibnamefont {Tsumura}}, \
  and\ \bibinfo {author} {\bibfnamefont {T.}~\bibnamefont {Kunihiro}},\ }\href
  {\doibase 10.1103/PhysRevC.92.064909} {\bibfield  {journal} {\bibinfo
  {journal} {Phys. Rev. C}\ }\textbf {\bibinfo {volume} {92}},\ \bibinfo
  {pages} {064909} (\bibinfo {year} {2015})},\ \Eprint
  {http://arxiv.org/abs/1507.04894} {arXiv:1507.04894 [hep-ph]} \BibitemShut
  {NoStop}%
\bibitem [{\citenamefont {Fotakis}\ \emph {et~al.}(2022)\citenamefont
  {Fotakis}, \citenamefont {Moln\'ar}, \citenamefont {Niemi}, \citenamefont
  {Greiner},\ and\ \citenamefont {Rischke}}]{Fotakis:2022usk}%
  \BibitemOpen
  \bibfield  {author} {\bibinfo {author} {\bibfnamefont {J.~A.}\ \bibnamefont
  {Fotakis}}, \bibinfo {author} {\bibfnamefont {E.}~\bibnamefont {Moln\'ar}},
  \bibinfo {author} {\bibfnamefont {H.}~\bibnamefont {Niemi}}, \bibinfo
  {author} {\bibfnamefont {C.}~\bibnamefont {Greiner}}, \ and\ \bibinfo
  {author} {\bibfnamefont {D.~H.}\ \bibnamefont {Rischke}},\ }\href {\doibase
  10.1103/PhysRevD.106.036009} {\bibfield  {journal} {\bibinfo  {journal}
  {Phys. Rev. D}\ }\textbf {\bibinfo {volume} {106}},\ \bibinfo {pages}
  {036009} (\bibinfo {year} {2022})},\ \Eprint
  {http://arxiv.org/abs/2203.11549} {arXiv:2203.11549 [nucl-th]} \BibitemShut
  {NoStop}%
\bibitem [{\citenamefont {Hu}\ and\ \citenamefont {Shi}(2022)}]{Hu:2022vph}%
  \BibitemOpen
  \bibfield  {author} {\bibinfo {author} {\bibfnamefont {J.}~\bibnamefont
  {Hu}}\ and\ \bibinfo {author} {\bibfnamefont {S.}~\bibnamefont {Shi}},\
  }\href {\doibase 10.1103/PhysRevD.106.014007} {\bibfield  {journal} {\bibinfo
   {journal} {Phys. Rev. D}\ }\textbf {\bibinfo {volume} {106}},\ \bibinfo
  {pages} {014007} (\bibinfo {year} {2022})},\ \Eprint
  {http://arxiv.org/abs/2204.10100} {arXiv:2204.10100 [hep-ph]} \BibitemShut
  {NoStop}%
\bibitem [{\citenamefont {Onsager}(1931{\natexlab{a}})}]{Onsager1931a}%
  \BibitemOpen
  \bibfield  {author} {\bibinfo {author} {\bibfnamefont {L.}~\bibnamefont
  {Onsager}},\ }\href {\doibase 10.1103/PhysRev.37.405} {\bibfield  {journal}
  {\bibinfo  {journal} {Phys. Rev.}\ }\textbf {\bibinfo {volume} {37}},\
  \bibinfo {pages} {405} (\bibinfo {year} {1931}{\natexlab{a}})}\BibitemShut
  {NoStop}%
\bibitem [{\citenamefont {Onsager}(1931{\natexlab{b}})}]{Onsager1931b}%
  \BibitemOpen
  \bibfield  {author} {\bibinfo {author} {\bibfnamefont {L.}~\bibnamefont
  {Onsager}},\ }\href {\doibase 10.1103/PhysRev.38.2265} {\bibfield  {journal}
  {\bibinfo  {journal} {Phys. Rev.}\ }\textbf {\bibinfo {volume} {38}},\
  \bibinfo {pages} {2265} (\bibinfo {year} {1931}{\natexlab{b}})}\BibitemShut
  {NoStop}%
\bibitem [{\citenamefont {Weil}\ \emph {et~al.}(2016)\citenamefont {Weil} \emph
  {et~al.}}]{SMASH:2016zqf}%
  \BibitemOpen
  \bibfield  {author} {\bibinfo {author} {\bibfnamefont {J.}~\bibnamefont
  {Weil}} \emph {et~al.} (\bibinfo {collaboration} {SMASH}),\ }\href {\doibase
  10.1103/PhysRevC.94.054905} {\bibfield  {journal} {\bibinfo  {journal} {Phys.
  Rev. C}\ }\textbf {\bibinfo {volume} {94}},\ \bibinfo {pages} {054905}
  (\bibinfo {year} {2016})},\ \Eprint {http://arxiv.org/abs/1606.06642}
  {arXiv:1606.06642 [nucl-th]} \BibitemShut {NoStop}%
\bibitem [{\citenamefont {Bala}\ \emph {et~al.}(2023)\citenamefont {Bala},
  \citenamefont {Kaczmarek}, \citenamefont {Lowdon}, \citenamefont
  {Philipsen},\ and\ \citenamefont {Ueding}}]{Bala:2023iqu}%
  \BibitemOpen
  \bibfield  {author} {\bibinfo {author} {\bibfnamefont {D.}~\bibnamefont
  {Bala}}, \bibinfo {author} {\bibfnamefont {O.}~\bibnamefont {Kaczmarek}},
  \bibinfo {author} {\bibfnamefont {P.}~\bibnamefont {Lowdon}}, \bibinfo
  {author} {\bibfnamefont {O.}~\bibnamefont {Philipsen}}, \ and\ \bibinfo
  {author} {\bibfnamefont {T.}~\bibnamefont {Ueding}},\ }\href@noop {} {\
  (\bibinfo {year} {2023})},\ \Eprint {http://arxiv.org/abs/2310.13476}
  {arXiv:2310.13476 [hep-lat]} \BibitemShut {NoStop}%
\end{thebibliography}%
	
\end{document}